\documentclass[conference]{IEEEtran}
\IEEEoverridecommandlockouts
\usepackage{cite}
\usepackage{amsmath,amssymb,amsfonts}
\usepackage{algorithmic}
\usepackage{graphicx}
\usepackage{textcomp}
\usepackage{xcolor}
\usepackage{multirow}
\usepackage{url}
\usepackage{breqn}
\usepackage{fancyhdr}
\def\BibTeX{{\rm B\kern-.05em{\sc i\kern-.025em b}\kern-.08em
    T\kern-.1667em\lower.7ex\hbox{E}\kern-.125emX}}

\fancypagestyle{firstpage}{
  \fancyhf{}
  
  \fancyfoot[L]{\footnotesize © 2023 IEEE. Personal use of this material is permitted. Permission from IEEE must be obtained for all other uses, in any current or future media, including reprinting/republishing this material for advertising or promotional purposes, creating new collective works, for resale or redistribution to servers or lists, or reuse of any copyrighted component of this work in other works. DOI: 10.1109/ICoDSE59534.2023.10291842}
}

\begin{document}

\title{Automated Chest X-Ray Report Generator Using Multi-Model Deep Learning Approach}

\author{\IEEEauthorblockN{1\textsuperscript{st} Arief Purnama Muharram}
\IEEEauthorblockA{\textit{School of Electrical Engineering and Informatics} \\
\textit{Institut Teknologi Bandung}\\
Bandung, Indonesia \\
23521013@std.stei.itb.ac.id}
\and
\IEEEauthorblockN{2\textsuperscript{nd} Hollyana Puteri Haryono}
\IEEEauthorblockA{\textit{School of Electrical Engineering and Informatics} \\
\textit{Institut Teknologi Bandung}\\
Bandung, Indonesia \\
23522013@std.stei.itb.ac.id}
\and
\IEEEauthorblockN{3\textsuperscript{rd} Abassi Haji Juma}
\IEEEauthorblockA{\textit{School of Electrical Engineering and Informatics} \\
\textit{Institut Teknologi Bandung}\\
Bandung, Indonesia \\
23522701@std.stei.itb.ac.id}
\and
\IEEEauthorblockN{4\textsuperscript{th} Ira Puspasari}
\IEEEauthorblockA{\textit{Computer Engineering} \\
\textit{Universitas Dinamika}\\
Surabaya, Indonesia\\
ira@dinamika.ac.id}
\and
\IEEEauthorblockN{5\textsuperscript{th} Nugraha Priya Utama\textsuperscript{*} 
\IEEEauthorblockA{\textit{School of Electrical Engineering and Informatics} \\
\textit{Institut Teknologi Bandung}\\
Bandung, Indonesia \\
utama@informatika.org \\ \thanks{\textsuperscript{*}Corresponding author}}}
}

\maketitle

\thispagestyle{firstpage}

\begin{abstract} 
Reading and interpreting chest X-ray images is one of the most radiologist's routines. However, it still can be challenging, even for the most experienced ones. Therefore, we proposed a multi-model deep learning-based automated chest X-ray report generator system designed to assist radiologists in their work. The basic idea of the proposed system is by utilizing multi binary-classification models for detecting multi abnormalities, with each model responsible for detecting one abnormality, in a single image. In this study, we limited the radiology abnormalities detection to only cardiomegaly, lung effusion, and consolidation. The system generates a radiology report by performing the following three steps: image pre-processing, utilizing deep learning models to detect abnormalities, and producing a report. The aim of the image pre-processing step is to standardize the input by scaling it to 128x128 pixels and slicing it into three segments, which covers the upper, lower, and middle parts of the lung. After pre-processing, each corresponding model classifies the image, resulting in a 0 (zero) for no abnormality detected and a 1 (one) for the presence of an abnormality. The prediction outputs of each model are then concatenated to form a 'result code'. The 'result code' is used to construct a report by selecting the appropriate pre-determined sentence for each detected abnormality in the report generation step. The proposed system is expected to reduce the workload of radiologists and increase the accuracy of chest X-ray diagnosis.
\end{abstract}


\section{Introduction}
Deep learning for image processing has rapidly advanced in recent years \cite{lecun15}, demonstrating promising results across a variety of domains, including radiology \cite{thrall18}. These kind of advancements have enabled improvements in radiological diagnosis and patient care\cite{thrall18}. Therefore, there is a demand for automated systems that can accurately and quickly process the growing volume of medical images \cite{thrall18}. Chest X-ray (CXR) imaging, in particular, is one of the most often utilized radiologic diagnostic tools, although its interpretation can be difficult even for the experienced radiologists\cite{lee13}. Radiologists can benefit from automated CXR report generation since it allows them to focus on more difficult cases while reducing the possibility of errors.

In this study, we proposed a multi-model deep learning-based automated CXR report generator system to help radiologists’ work better and faster. The proposed system is designed to detect abnormalities in a CXR image and then produce the corresponding report. Our approach was motivated by previous research that successfully applied deep learning algorithms to the processing of chest radiographs \cite{rajkomar17,cicero17,lakhani17,saha20,deb22,tang20,bhatt21}. However, most of the studies mentioned have primarily focused on detecting or classifying a single abnormality in the image. Therefore, in this study, we aimed to detect multiple abnormalities in radiology images. We hypothesized that using multi-model, with each model responsible for an abnormal finding, can detect multiple abnormalities present within a single image.

This paper will be structured as follows: Firstly, we will discuss related work on deep learning in radiology and mention several previous research studies in the field. Next, we will present the methodology and technical aspects of our suggested system. In the Results section, we will describe and discuss our experimental findings. Finally, we will conclude our study.

\begin{table*}[ht]
\caption{Literature Review}
\centering
\begin{tabular}{|p{1cm}|l|p{4cm}|p{4cm}|p{4cm}|} \hline
\multicolumn{1}{|c|}{\textbf{Author}} & \multicolumn{1}{c|}{\textbf{Year}} & \multicolumn{1}{c|}{\textbf{Aim}} & \multicolumn{1}{c|}{\textbf{Algorithm Used}} & \textbf{Performance} \\ \hline
Rajkomar et al. \cite{rajkomar17} & 2017 & Developing a deep convolutional neural network for classifying the view orientations of chest radiographs. & GoogLeNet & Accuracy of 100\% and 38 image classifications per second. \\ \hline

Lakhani et al. \cite{lakhani17}& 2017 & Developing CNNs for automated classification of pulmonary tuberculosis using chest radiograph & AlexNet, GoogLeNet & The best performance is achieved by the ensemble of AlexNet and GoogLeNet, with an Area Under the Curve (AUC) of 0.99. \\ \hline

Cicero et al. \cite{cicero17} & 2017 & Developing and test a deep convolutional neural network for frontal chest radiograph abnormality classification and computer-aided detection & GoogLeNet & The sensitivity, specificity, and AUC for pleural effusion were 0.91, 0.91, and 0.962, respectively; for pulmonary edema, they were 0.82, 0.82, and 0.868; for consolidation, 0.74, 0.75, and 0.850; for cardiomegaly, 0.81, 0.80, and 0.875; and for pneumothorax, 0.78, 0.78, and 0.861. \\ \hline

Saha et al. \cite{saha20} & 2020 & Developing multi model based ensemble to detect COVID-19 from CXR Images & Multiple CNN models (CNNs, MobileNet, InceptionV3, DenseNet201, DenseNet121 and Xception) ensembled using feature concatenation and decision fusion & Accuracy of 96\% using decision fusion scheme for 3 class classification and 89.21\% for 4 class classification. Accuracy of 95.84\% using feature concatenation scheme for 3 class classification and 89.26\% for 4 class classification \\ \hline

Deb et al. \cite{deb22} & 2022 & Developing a multi model ensemble based DCNN structure & VGGNet, GoogLeNet, DenseNet, and NASNet & An accuracy of 88.98\% was achieved for the three-class classification, and a 98.58\% accuracy was attained for the binary class classification using the publicly available dataset. \\ \hline

Tang et al. \cite{tang20} & 2020 & Developing deep CNN model to classify normal vs abnormal CXRs & AlexNet, VGG, GoogLeNet, ResNet, and DenseNet & ROC score for ResNet18, ResNet50, GoogLeNet, and DenseNet are by around 98\% using both NIH CXR and WCMC pediatric dataset. \\ \hline

Bhatt et al. \cite{bhatt21} & 2021 & Developing classification model for pulmonary consolidations detection in CXR images using  transfer learning and progressive resizing techniques & VGG-19 (baseline for binary classification) and EfficientNet-B3 (baseline for multi-class classification) & Achieved 100\% accuracy \\ \hline

Amjoud et al. \cite{amjoud21} & 2021 & Developing automated report generation for CXR using Transformer-based model & Transfer learning and transformer-based approach & Achieved BLUE-1, BLUE-2, and ROUGE metrics, where the scores were 0.479, 0.359, and 0.380, respectively. \\ \hline

Ghadekar et al. \cite{ghadekar21} & 2021 & Developing report generation using VGG for chest diseases diagnosis and encoder-decoder model for report generation. & VGG16 and encoder-decoder model & Accuracy of 88\% for classification images and 96\% for report generation. \\ \hline

\end{tabular}
\label{table:literature_review}
\end{table*}

\section{Related work}
In recent years, medical image analysis has emerged as a major use of deep learning, with the potential to aid radiologists in diagnosing a disease. Deep learning algorithms, in particular, have shown promise in the identification and classification of abnormalities on chest radiographs. One of the popular and widely-used deep learning algorithms in the field is Convolutional Neural Network (CNN).

LeCun et al. \cite{lecun15} present an in-depth overview of deep learning, including CNNs, which have been demonstrated to be successful for image classification tasks. CNN is a form of artificial neural network (ANN) that is extensively employed in image recognition and processing applications. It is based on the structure and function of the visual cortex in animals, which is responsible for visual information processing \cite{lecun15}.

In a CNN, input data is fed into a sequence of convolutional layers, which extract features from the input image by applying a set of filters. To inject non-linearity into the model, the output of each convolutional layer is then processed through a non-linear activation function such as ReLU (Rectified Linear Unit). As a result, the input data is represented by a set of high-level features that become increasingly complex and abstract.

The output of the convolutional layers is often sent into one or more fully connected layers, which conduct classification or regression based on the retrieved features. The network's ultimate output is a probability distribution over the various classes or values of the output variable.

CNNs have proven to be extremely effective in image classification, object detection, and segmentation tasks, and have been utilized in a variety of applications including as self-driving cars, medical diagnostics, and facial recognition.

CNNs were shown by Rajkomar et al. \cite{rajkomar17} to be capable of classifying view orientations of chest radiographs with excellent accuracy. Lakhani et al. \cite{lakhani17} used chest radiographs to construct a CNN-based model for the automated classification of pulmonary tuberculosis, obtaining high performance and indicating the promise for deep learning in the disease detection. Cicero et al. \cite{cicero17} used a CNN to detect and classify abnormalities on chest radiographs, with good sensitivity and specificity. Tang et al. \cite{tang20} compare several existing CNN architectures such as AlexNet, VGGNet, ResNet, GoogLeNet, and DenseNet with tested on NIH CXR dataset using ROC score for evaluation. It found that DenseNet, GoogLeNet, and ResNet has performance of ROC score more than 98\%. Bhatt et al. \cite{bhatt21} use modified VGG-19 and EfficientNet-B3 architecture and applied transfer learning technique to identify pneumonia and COVID-19 in CXR images. Table \ref{table:literature_review} describes in detail previous research on deep learning in radiology.

There are several papers that used multi-model based ensembling approaches for detection of COVID19 from CXR images. Saha et al. \cite{saha20} use ensembling approaches with both feature concatenation and decision fusion technique. Deb et al. \cite{deb22} develop a multi-model ensemble based on DCNN architecture with combining VGGnet, GoogLeNet, DenseNet, and NASNet.

Report Generation of CXR using deep learning approach has been done before by Amjoud et al. \cite{amjoud21} and Ghadekar et al. \cite{ghadekar21}. Encoder-Decoder approach has been the focus in Ghadekar et al. \cite{ghadekar21} and giving a results of 96\% of accuracy. However, they do not mentioned the encoder-decoder architecture used, nor do they explain the characteristics of the generated report. Amjoud et al. \cite{amjoud21} focused on combining CNN for CXR image classification and transformers for report generation.

Compared to the mentioned studies, our proposed method utilizes a multi-model approach to detect multiple abnormalities present in the image. The results from each model are then used to generate the report. We argue that using multiple models, each dedicated to a specific abnormality, leads to more 'responsible' results.

\section{Methodology}

\subsection{System Design}

\begin{figure}[hb]
\centerline{\includegraphics[width=0.48\textwidth]{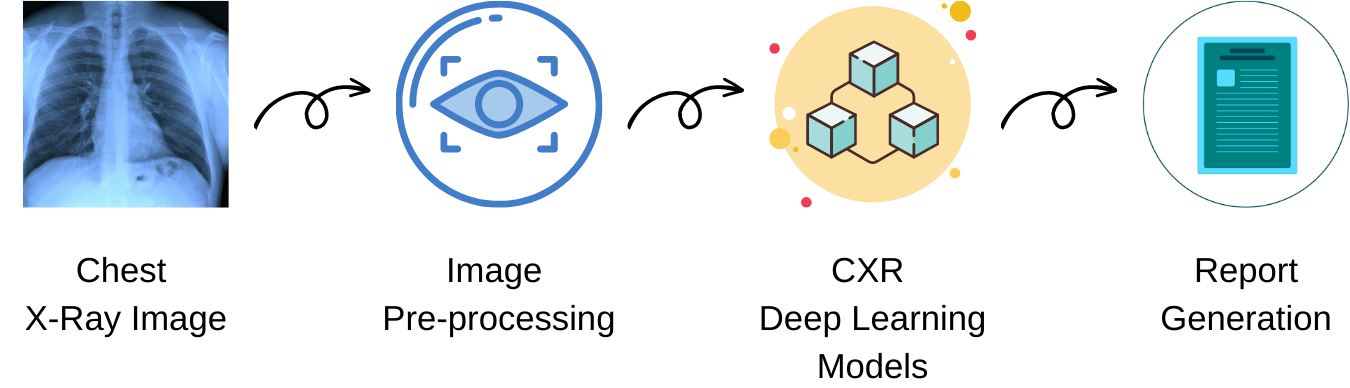}}
\caption{Three fundamental steps in our proposed methodology}
\label{fig:proposed_methodology}
\end{figure}

Based on our literature review, we are motivated to create a deep-learning model for quantifying and localizing abnormalities on chest radiographs and generating corresponding X-ray reports, to support radiologists in diagnosing a disease. Our proposed methodology consists of three fundamental steps: image pre-processing, CXR deep learning models, and report production (Figure \ref{fig:proposed_methodology}).

\begin{figure}[ht]
\centerline{\includegraphics[width=0.45\textwidth]{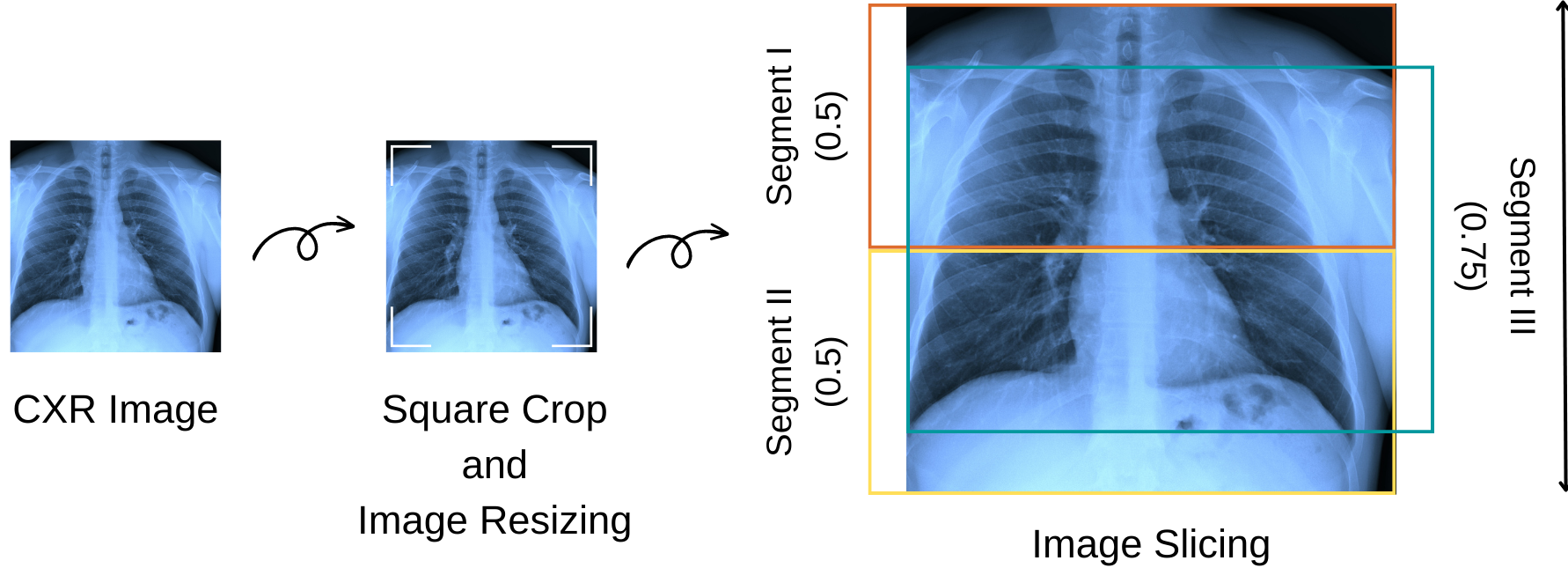}}
\caption{Image pre-processing step}
\label{fig:image_preprocessing_step}
\end{figure}

Image pre-processing is the first step of the entire process, as illustrated in Figure \ref{fig:image_preprocessing_step}. In this step, each image is resized to 128x128 and square-cropped to ensure consistent dimensionality. Then, the image is divided into three segments (Segment I, II, and III) to reduce the amount of "information" passed to the deep learning model in the subsequent step. Segment I covers the upper part of the lung, segment II covers the lower part of the lung, and segment III covers the middle part of the lung.

\begin{figure}[ht]
\centerline{\includegraphics[width=0.45\textwidth]{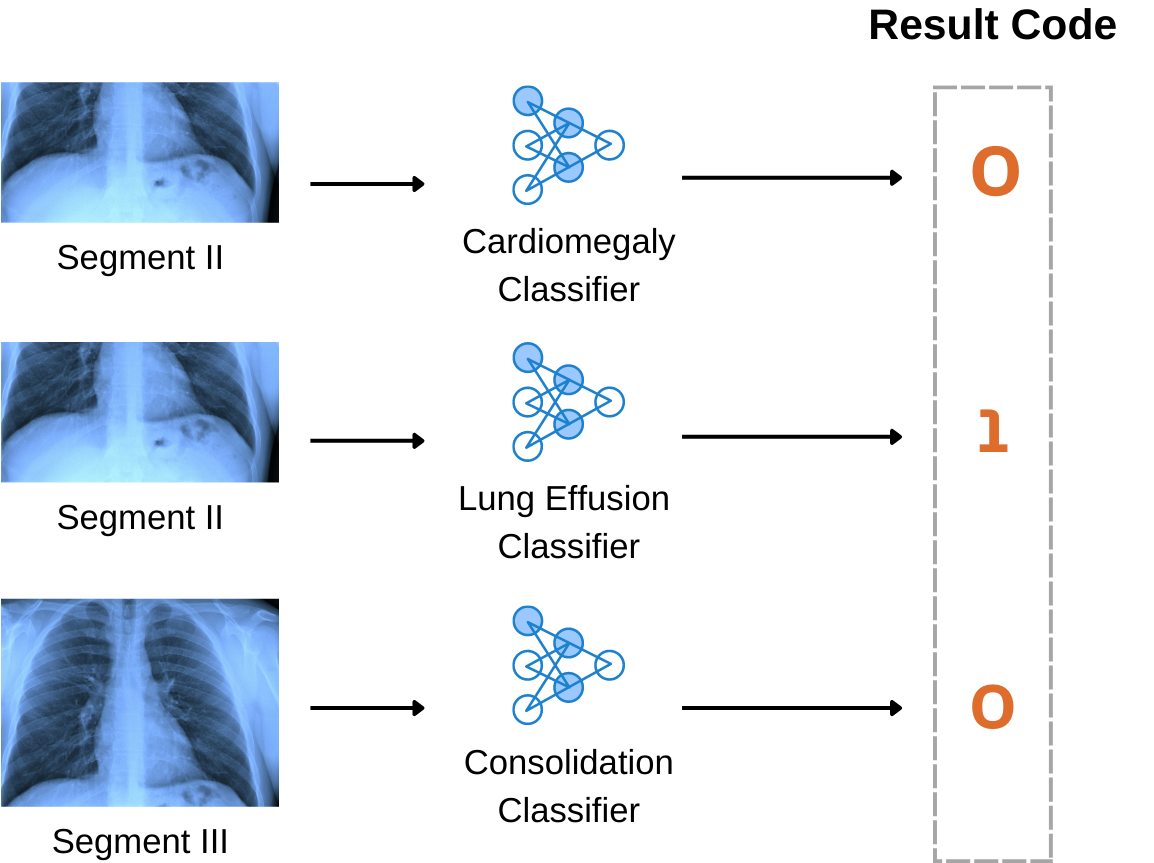}}
\caption{Multi-models breakdown and result code}
\label{fig:deep_learning_modules}
\end{figure}

The second step involves training deep learning models to detect each abnormality. Three models were trained, one for detecting cardiomegaly, another for detecting lung effusion, and the third for detecting consolidation. However, instead of using the whole image as input for the model, we used the corresponding sliced image based on the specific abnormality of interest (Figure \ref{fig:deep_learning_modules}). For example, we used segment II as the input for the cardiomegaly and lung effusion models because these abnormalities are more likely to be found in the lower part of the lungs, while segment III was used as input for the consolidation model because this abnormality is more likely to be found in the middle part of the lungs. The NIH CXR dataset was used to train the models \cite{wang17}. For the experiments, we used the "one-factor-at-a-time" strategy, using the learning rate, optimizer, and pre-trained models as the observable factors. In addition, we compared the ResNet18 \cite{zhang16}, ResNet50 \cite{zhang16}, and GoogleLeNet \cite{szegedy15} pre-trained models, which are popular CNN algorithms. 

We trained our models on a high-performance computing server with an Intel Xeon Silver 4208 CPU 3.2~GHz, 256~GB of RAM, and an NVIDIA Quadro RTX 5000 GPU with 16~GB of RAM. For our deep learning library, we used Torch version 1.13.1 and Torchvision version 0.14.1 in a Python 3.9.6 environment, and for deployment, we used Gradio version 3.28.3.

\begin{figure}[ht]
\centerline{\includegraphics[width=0.48\textwidth]{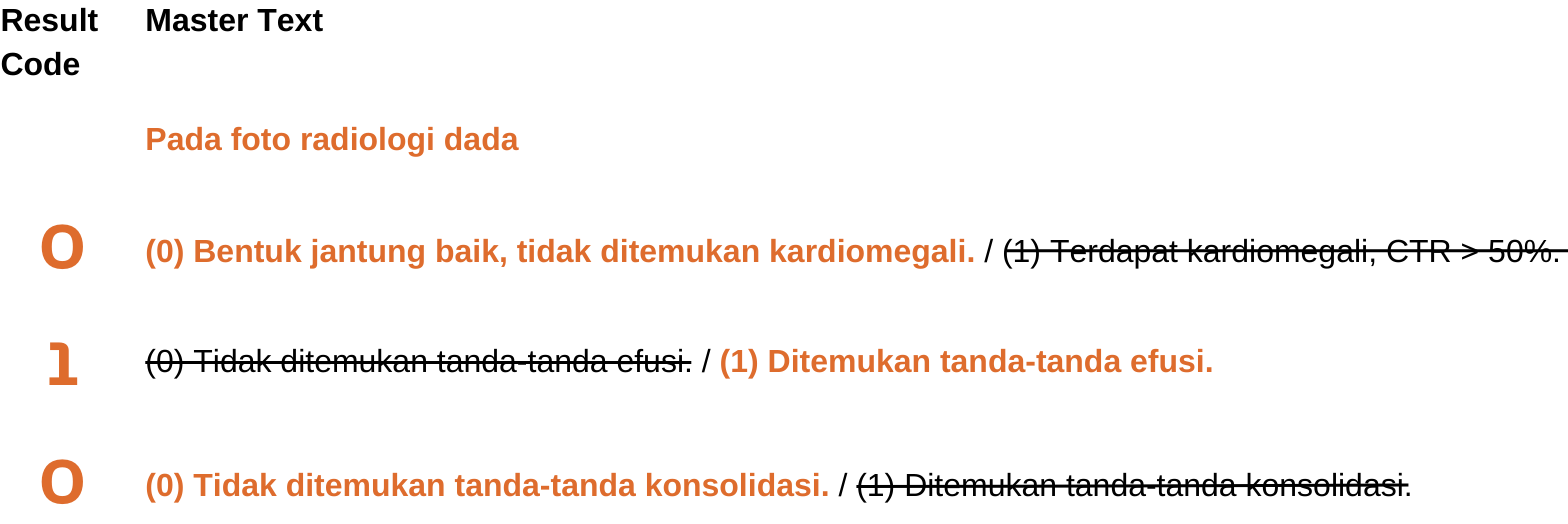}}
\caption{ Report generation is based on the result code and master text}
\label{fig:report_generation}
\end{figure}

Finally, the last step is report generation. The purpose of this step is to map the result code, which is an aggregation of each model’s prediction output, to the corresponding sentence in the master text (Figure \ref{fig:report_generation}). For example, if a cardiomegaly abnormality is detected (coded as “1”), the sentence “Terdapat kardiomegali, CTR $<$ 50\%” will be selected from the master text. Otherwise, the sentence “Bentuk jantung baik, tidak ditemukan kardiomegali” will be used.

\subsection{Evaluation Strategy}
We used accuracy as our evaluation metric, both for individual trained models and for overall system performance. Accuracy is determined by calculating the total number of correct predictions divided by the total number of predictions. However, when assessing the system's performance, a true prediction is only achieved when all constituent models provide accurate predictions. For instance, if the prediction labels for the cardiomegaly, lung effusion, and consolidation models are [0, 0, 1], respectively, and the ground truth labels are [0, 0, 0], then the prediction result is considered incorrect.

\subsection{Problem Limitations}
There are several limitations in our study that should be acknowledged, including:

\begin{itemize}
\item Radiology image limitation: The suggested method was trained using adult frontal CXRs and is only intended to be used for adult patients' frontal CXR images. Therefore, the system may not be suitable for interpreting X-rays of pediatric patients.

\item Abnormalities detection limitation: The method is designed to identify specific abnormalities, namely cardiomegaly, lung effusion, and consolidation. As a result, any additional abnormalities present in the image that fall outside the scope of this study are not be detected by the system.

\end{itemize}

\section{Result}

\subsection{Dataset Pre-processing}
We used The NIH CXR Dataset \cite{wang17} to train our models, which comprises hundreds of thousands of CXR images with 14 different labels for abnormality findings. However, we only used 3 of the 14 abnormality labels, namely cardiomegaly, lung effusion, and consolidation. To ensure data quality, we employed domain experts to conduct reannotation. For each abnormality, we selected 2,000 images at random (1,000 images with the presence of abnormality and 1,000 without). Any images that did not meet our criteria, such as low image quality, incorrect orientation, or misdiagnosis, were excluded. Following annotation, the dataset was randomly split into the training (70\%) and testing (30\%) datasets. Table \ref{table:dataset} presents the results of our annotated dataset.

\begin{table}[hb]
\caption{Total number of images in the reannotated dataset}
\centering
\begin{tabular}{|lccc|} \hline
\multicolumn{1}{|c|}{\multirow{2}{*}{\textbf{Abnormalities}}} & \multicolumn{2}{c|}{\textbf{Labels}} & \multicolumn{1}{c|}{\multirow{2}{*}{\textbf{Total (N)}}} \\ \cline{2-3}
\multicolumn{1}{|c|}{} & \multicolumn{1}{c|}{\textbf{False (n)}} & \multicolumn{1}{c|}{\textbf{True (n)}} & \multicolumn{1}{|c|}{} \\ 
\hline
\multicolumn{1}{|l|}{Cardiomegaly} & \multicolumn{1}{c|}{615} & \multicolumn{1}{c|}{793} & \multicolumn{1}{c|}{1408} \\ \hline
\multicolumn{1}{|l|}{Lung Effusion} & \multicolumn{1}{c|}{658} & \multicolumn{1}{c|}{896} & \multicolumn{1}{c|}{1554} \\ \hline
\multicolumn{1}{|l|}{Consolidation} & \multicolumn{1}{c|}{454} & \multicolumn{1}{c|}{652} & \multicolumn{1}{c|}{1106} \\ \hline
\end{tabular}
\label{table:dataset}
\end{table}

To evaluate the system's performance, we created a separate dataset, named the "system evaluation dataset," by randomly selecting 200 images from The NIH CXR dataset. In contrast to the model training dataset, this dataset comprises three labels: cardiomegaly, lung effusion, and consolidation, each with a binary value of 0 for absence or 1 for presence of abnormality. This dataset is designed to assess the system's ability to accurately predict all three labels simultaneously. Additionally, domain experts have reannotated the dataset to ensure its quality, just as we did with the model training dataset.

\subsection{Modeling}

There were several parameters that we considered for hyperparameter tuning, such as learning rate, optimizer, and pretrained model. Our hyperparameter tuning strategy was based on the "one-factor-at-a-time" approach, where we varied one parameter at a time while keeping the others at a default value. For each parameter value, we compared the resulting training and testing accuracy. Hyperparameter tuning was performed for each model, and the results are presented in Tables \ref{table:hyperparameter_tuning_result_cardiomegaly}, \ref{table:hyperparameter_tuning_result_effusion}, and \ref{table:hyperparameter_tuning_result_consolidation}.

After completing hyperparameter tuning, we found that the optimal values were different for each abnormality. The most optimal learning rates were 1e-3 for effusion and cardiomegaly, and 1e-4 for consolidation. Adam was the best optimizer since we found that SGD did not produce the best accuracy. Also, the most optimal pre-trained model for each abnormalities were different, such as GoogLeNet for cardiomegaly, ResNet50 for effusion, and ResNet18 for consolidation. Using these optimal parameters for each abnormality, we trained the final model and recorded the training and testing accuracy as shown on Table \ref{table:final_model_result}. The final models will be the models used in the system.

\begin{table}[ht]
\caption{Hyperparameter Tuning Result for Cardiomegaly}
\begin{tabular}{|l|l|l|rr|}
\hline
\multirow{2}{*}{\textbf{Step}} & \multirow{2}{*}{\textbf{Hyperparameter}} & \multicolumn{1}{c|}{\multirow{2}{*}{\textbf{Value}}} & \multicolumn{2}{c|}{\textbf{Accuracy}} \\ \cline{4-5} 
 &  & \multicolumn{1}{c|}{} & \multicolumn{1}{l|}{\textbf{Training}} & \multicolumn{1}{l|}{\textbf{Testing}} \\ \hline
\multirow{4}{*}{1} & \multirow{4}{*}{Learning Rate} & 1e-2 & \multicolumn{1}{r|}{75.4\%} & 80.5\% \\ \cline{3-5} 
 &  & \textbf{1e-3} (default) & \multicolumn{1}{r|}{\textbf{87.1\%}} & \textbf{87\%} \\ \cline{3-5} 
 &  & 1e-4 & \multicolumn{1}{r|}{82.6\%} & 81.5\% \\ \cline{3-5} 
 &  & 1e-5 & \multicolumn{1}{r|}{78.1\%} & 79.8\% \\ \hline
\multirow{2}{*}{2} & \multirow{2}{*}{Optimizer} & \textbf{Adam} (default) & \multicolumn{1}{r|}{\textbf{87.1\%}} & \textbf{87\%} \\ \cline{3-5} 
 &  & SGD & \multicolumn{1}{r|}{72.4\%} & 80.5\% \\ \hline
\multirow{3}{*}{3} & \multirow{3}{*}{Pre-trained Model} & ResNet50 (default) & \multicolumn{1}{r|}{87.1\%} & 87\% \\ \cline{3-5} 
 &  & ResNet18 & \multicolumn{1}{r|}{85\%} & 90.3\% \\ \cline{3-5} 
 &  & \textbf{GoogLeNet} & \multicolumn{1}{r|}{\textbf{90.1\%}} & \textbf{90.1\%} \\ \hline
\end{tabular}
\label{table:hyperparameter_tuning_result_cardiomegaly}
\end{table}

\begin{table}[ht]
\caption{Hyperparameter Tuning Result for Lung Effusion}
\begin{tabular}{|l|l|l|rr|}
\hline
\multirow{2}{*}{\textbf{Step}} & \multirow{2}{*}{\textbf{Hyperparameter}} & \multicolumn{1}{c|}{\multirow{2}{*}{\textbf{Value}}} & \multicolumn{2}{c|}{\textbf{Accuracy}} \\ \cline{4-5} 
 &  & \multicolumn{1}{c|}{} & \multicolumn{1}{l|}{\textbf{Training}} & \multicolumn{1}{l|}{\textbf{Testing}} \\ \hline
\multirow{4}{*}{1} & \multirow{4}{*}{Learning Rate} & 1e-2 & \multicolumn{1}{r|}{77.5\%} & 78.4\% \\ \cline{3-5} 
 &  & \textbf{1e-3} (default) & \multicolumn{1}{r|}{\textbf{90.2\%}} & \textbf{91.8\%} \\ \cline{3-5} 
 &  & 1e-4 & \multicolumn{1}{r|}{87\%} & 87\% \\ \cline{3-5} 
 &  & 1e-5 & \multicolumn{1}{r|}{82\%} & 84.6\% \\ \hline
\multirow{2}{*}{2} & \multirow{2}{*}{Optimizer} & \textbf{Adam} (default) & \multicolumn{1}{r|}{\textbf{90.2\%}} & \textbf{91.8\%} \\ \cline{3-5} 
 &  & SGD & \multicolumn{1}{r|}{76.4\%} & 79.4\% \\ \hline
\multirow{3}{*}{3} & \multirow{3}{*}{Pre-trained Model} & \textbf{ResNet50} (default) & \multicolumn{1}{r|}{\textbf{90.2\%}} & \textbf{91.8\%} \\ \cline{3-5} 
 &  & ResNet18 & \multicolumn{1}{r|}{91.5\%} & 87.4\% \\ \cline{3-5} 
 &  & GoogLeNet & \multicolumn{1}{r|}{87.8\%} & 89.4\% \\ \hline
\end{tabular}
\label{table:hyperparameter_tuning_result_effusion}
\end{table}

\begin{table}[ht]
\caption{Hyperparameter Tuning Result for Consolidation}
\begin{tabular}{|l|l|l|rr|}
\hline
\multirow{2}{*}{\textbf{Step}} & \multirow{2}{*}{\textbf{Hyperparameter}} & \multicolumn{1}{c|}{\multirow{2}{*}{\textbf{Value}}} & \multicolumn{2}{c|}{\textbf{Accuracy}} \\ \cline{4-5} 
 &  & \multicolumn{1}{c|}{} & \multicolumn{1}{l|}{\textbf{Training}} & \multicolumn{1}{l|}{\textbf{Testing}} \\ \hline
\multirow{4}{*}{1} & \multirow{4}{*}{Learning Rate} & 1e-2 & \multicolumn{1}{r|}{67.5\%} & 68.9\% \\ \cline{3-5} 
 &  & 1e-3 (default) & \multicolumn{1}{r|}{79.1\%} & 78\% \\ \cline{3-5} 
 &  & \textbf{1e-4} & \multicolumn{1}{r|}{\textbf{79.1\%}} & \textbf{81\%} \\ \cline{3-5} 
 &  & 1e-5 & \multicolumn{1}{r|}{69.6\%} & 74.7\% \\ \hline
\multirow{2}{*}{2} & \multirow{2}{*}{Optimizer} & \textbf{Adam} (default) & \multicolumn{1}{r|}{\textbf{79.1\%}} & \textbf{78\%} \\ \cline{3-5} 
 &  & SGD & \multicolumn{1}{r|}{66\%} & 65.6\% \\ \hline
\multirow{3}{*}{3} & \multirow{3}{*}{Pre-trained Model} & ResNet50 (default) & \multicolumn{1}{r|}{79.1\%} & 78\% \\ \cline{3-5} 
 &  & \textbf{ResNet18} & \multicolumn{1}{r|}{\textbf{82.7\%}} & \textbf{78.4\%} \\ \cline{3-5} 
 &  & GoogLeNet & \multicolumn{1}{r|}{78.5\%} & 73.3\% \\ \hline
\end{tabular}
\label{table:hyperparameter_tuning_result_consolidation}
\end{table}

\begin{table}[ht]
\caption{Result of each final model}
\centering
\begin{tabular}{|l|rr|}
\hline
\multicolumn{1}{|c|}{\multirow{2}{*}{\textbf{Abnormality}}} & \multicolumn{2}{c|}{\textbf{Accuracy}} \\ \cline{2-3} 
\multicolumn{1}{|c|}{} & \multicolumn{1}{c|}{\textbf{Training}} & \multicolumn{1}{c|}{\textbf{Testing}} \\ \hline
Cardiomegaly & \multicolumn{1}{r|}{89.2\%} & 88.7\% \\ \hline
Lung Effusion & \multicolumn{1}{r|}{88.6\%} & 86.1\% \\ \hline
Consolidation & \multicolumn{1}{r|}{75.9\%} & 81\% \\ \hline
\end{tabular}
\label{table:final_model_result}
\end{table}

\subsection{System Evaluation}

\begin{figure}[hb]
\centerline{\includegraphics[width=0.45\textwidth]{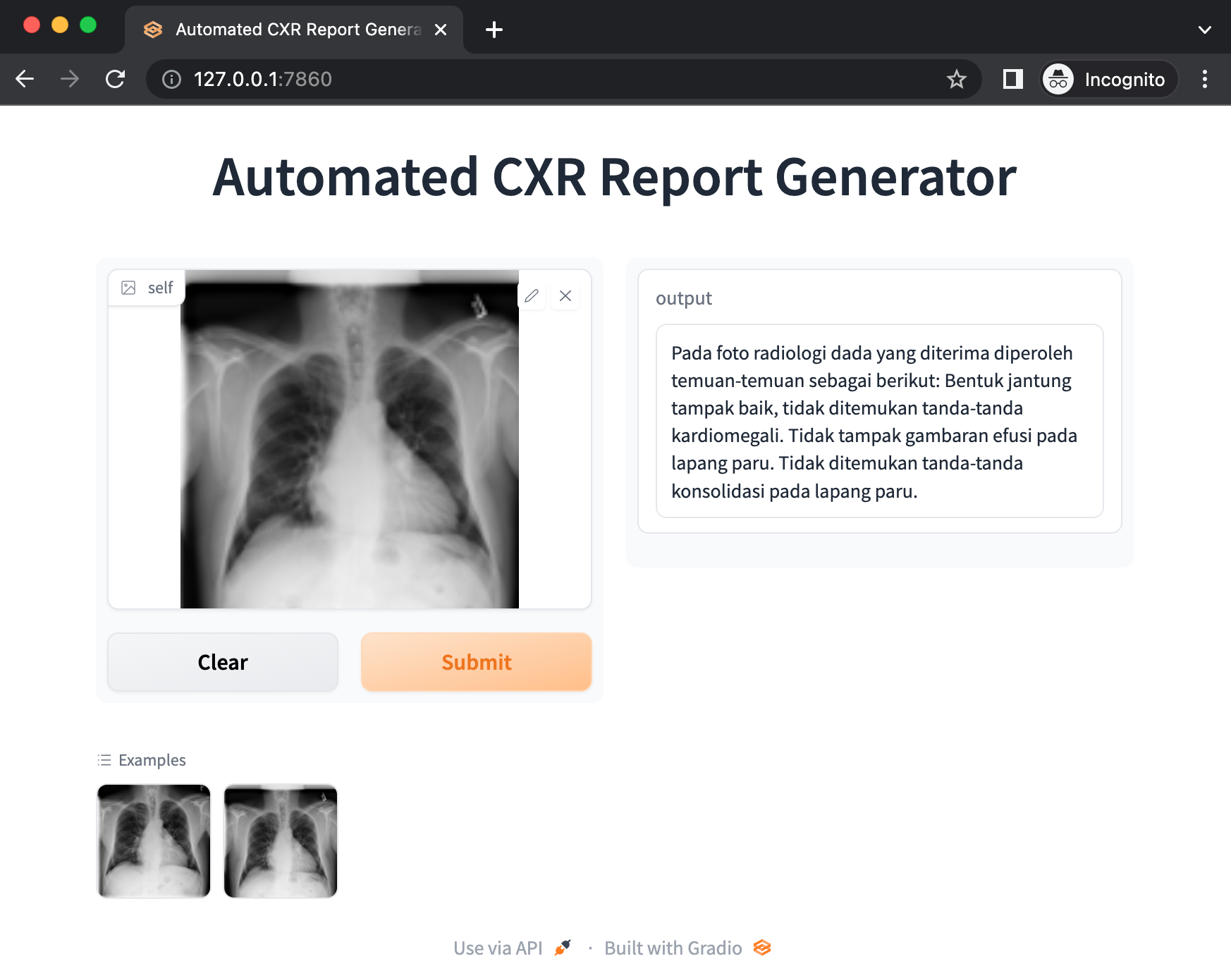}}
\caption{System example}
\label{fig:system_example}
\end{figure}

\begin{table}[ht!]
\caption{System Evaluation}
\centering
\begin{tabular}{|l|r|}
\hline
\multicolumn{1}{|c|}{\textbf{Abnormality}} & \multicolumn{1}{c|}{\textbf{Accuracy}} \\ \hline
Cardiomegaly & 52\% \\ \hline
Lung Effusion & 80\% \\ \hline
Consolidation & 40\% \\ \hline
\textbf{Cardiomegaly + Lung Effusion + Consolidation} & \textbf{20\%} \\ \hline
\end{tabular}
\label{table:system_evaluation}
\end{table}

We successfully created and deployed the system in the form of a web-based application (Figure \ref{fig:system_example}). Tthe system's performance evaluation was done using the system evaluation dataset. For a correct prediction, the system needed to accurately predict all three abnormalities; any incorrect prediction would render the overall result incorrect. Despite our hypothesis that the multi-model approach would yield promising results, the system's accuracy was only 20\% (Table \ref{table:system_evaluation}).

Our proposed system exhibits significantly lower performance in comparison to the studies conducted by Amjoud et al. \cite{amjoud21} and Ghadekar et al. \cite{ghadekar21} in the generation of radiology reports. To investigate the poor performance of the system, we conducted an error analysis that included a model breakdown analysis. We discovered that the accuracy of the cardiomegaly, lung effusion, and consolidation models towards the system evaluation dataset were only 52\%, 80\%, and 40\%, respectively (Table \ref{table:system_evaluation}). Of the three models, the worst performance results were obtained by the cardiomegaly and consolidation models.

The poor results obtained by the cardiomegaly and consolidation models indicate that these models failed to generalize well. Although the system evaluation dataset and model training dataset were sourced from the same dataset (the NIH CXR Dataset \cite{wang17}), this doesn't guarantee that the data is the "same". Medical images are highly heterogeneous, which can increase the risk of overfitting and lead to a loss of generalizability for the models \cite{thrall18}.

Furthermore, the errors of each model contribute to the poor performance of the system. Based on the accuracy, it is evident that the error rates of the cardiomegaly, lung effusion, and consolidation models were 48\% \eqref{eqn:error_pa}, 20\% \eqref{eqn:error_pb}, and 60\% \eqref{eqn:error_pc}, respectively. The combination of these errors could make it difficult for the system to make accurate predictions, and this could explain the poor performance results.

Below is our mathematical proof of probability to explain the poor performance of our system. Let $A$, $B$, and $C$ denote the cardiomegaly, lung effusion, and consolidation models, respectively. $P_{correct}(A)$, $P_{correct}(B)$, and $P_{correct}(C)$ represent the probabilities of accurately predicting the cardiomegaly, lung effusion, and consolidation models, respectively. In this case, the models' probabilities of accurate prediction are the models' accuracies themselves. Since the models function independently, we can assume that the events are also independent. We can calculate the probabilities of correct and error predictions using the following formula.

\begin{dmath}
P_{error}(A) = 1 - P_{correct}(A)
= 1 - 0.52
= 0.48
\label{eqn:error_pa}
\end{dmath}

\begin{dmath}
P_{error}(B) = 1 - P_{correct}(B)
= 1 - 0.8
= 0.2
\label{eqn:error_pb}
\end{dmath}

\begin{dmath}
P_{error}(C) = 1 - P_{correct}(C)
= 1 - 0.4
= 0.6
\label{eqn:error_pc}
\end{dmath}

\begin{dmath}
P_{correct}(A \cap B \cap C) = P_{correct}(A) \times P_{correct}(B) \times P_{correct}(C)
= 0.52 \times 0.8 \times 0.4
= 0.1664
\end{dmath}

\begin{dmath}
P_{error}(A \cup B \cup C) = P_{error}(A) + P_{error}(B) + P_{error}(C) - P_{error}(A \cap B) - P_{error}(A \cap C) - P_{error}(B \cap C) + P_{error}(A \cap B \cap C)
= 0.48 + 0.2 + 0.6 - (0.48 \times 0.2) - (0.48 \times 0.6) - (0.2 \times 0.6) + (0.48 \times 0.2 \times 0.6)
= 0.8336
\end{dmath}

The calculation shows that the probability of accurate predictions can be as low as $\approx 16\%$ while the probability of error predictions can be as high as $\approx 83\%$, which explains the system's poor performance in mathematically. Additionally, this highlights that the performance of the overall system is affected by the individual models' performance, which could explain the system's inferior performance when compared to the Amjoud et al. \cite{amjoud21} and Ghadekar et al. \cite{ghadekar21} studies.

\section{Conclusion}
We developed an automated CXR report generator based on multi-model deep learning, capable of detecting and classifying abnormalities in CXR images. Unfortunately, our system's overall accuracy was only 20\%, indicating that our multi-model technique of using separate models for each abnormality does not have the potential to significantly improve the accuracy and efficiency of radiological report generation.

Nonetheless, our method has a number of limitations that require further investigation in future research. One of the limitations is the dataset used in this study, which is unable to capture the heterogeneity of medical images, thereby limiting the model's ability to generalize the data. Consequently, the model's underperformance resulted in poor system performance. Future studies should aim to increase the amount of validated radiology data or develop alternative methods or algorithms to detect multiple abnormalities in a single radiology image.

\section*{Acknowledgment}
This work was the IF5200 Applied Research Project course project of the first, second, and third authors, under the supervision of the fourth and fifth authors. We would like to express our gratitude to the Informatics Master Study Program, School of Electrical Engineering and Informatics, Institut Teknologi Bandung, for providing us access to their high-performance computing server. Our research project would not have been possible without this vital resource. Furthermore, we also extend our appreciation to our colleague, Komang Shary Karismaputri, M.D., a radiology resident from Cipto Mangunkusumo Hospital at the Faculty of Medicine, Universitas Indonesia. Her insightful contributions to this research, particularly in providing valuable insights into the interpretation of radiology images, have greatly enriched the quality and depth of our study.

\end{document}